\documentclass[12pt]{article}

\input epsf
\usepackage{epsfig,graphicx,epsf}
\usepackage{amsmath,amsfonts,amssymb,amsbsy,hyperref}
\hypersetup{colorlinks=true,linkcolor=blue,citecolor=blue}

\def\i{\mbox{i}}
\def\\i{\mbox{\scriptsize{i}}}

\font\xcm = xcmitt10 scaled \magstep1

\title{pp elastic scattering at ISR and LHC energies}
\author{M. L. Nekrasov\\
{\small\it 
Institute for High Energy Physics, NRC ``Kurchatov
Institute'',}  \vspace*{-4\baselineskip}\\
{\small\it Protvino 142281, Russia} }
\date{}
\begin{document}
\maketitle

\begin{abstract}
We describe data on elastic $pp$ scattering at ISR and LHC energies using the analytical representation of the amplitude as the sum of three exponentials equipped with complex phases. In addition, we take into account the Coulomb and Coulomb-nuclear contributions in the entire transfer region and strong ``perturbative'' contributions in the large transfer region. Refined values of basic scattering characteristics are obtained. Significant changes in the behavior of the amplitude are detected at the transition from ISR to LHC energies. The results obtained indicate a two-layer structure of the colliding protons. In the transition to 13 TeV, a sharp increase in the size of the inner layer and its activity in the scattering is observed. A possible interpretation of this effect is discussed.
\end{abstract}

\section{Introduction}\label{sec1}

The main purpose of studying the diffraction scattering of protons is to reveal their structure and the features of strong interactions in the nonperturbative region. The usual way to deal with these problems involves considering different models to choose the one that best explains the data. A complementary way is to use data to reconstruct the scattering amplitude based on simple assumptions and search for an interpretation of its components. The second type of research is sometimes called ``model-independent''.

A prominent representative of the latter approach is the model of Barger and Phillips \cite{Phillips1973}. They proposed the amplitude as two exponentials with a relative phase, and based on this gave a description of the data in the ISR energy range and a limited transfer range 0.15 GeV$^2 < |t| < 5$ GeV$^2$. Ultimately, they interpreted the amplitude as the sum of the ``old'' Pomeron with equal-to-one intercept and a non-Regge background. 

Afterwards the same model was used \cite{Pancheri2012} to describe the preliminary LHC data at 7 TeV \cite{TOTEM2011a,TOTEM2011b}. However, it was later found \cite{Pancheri2013} that the final LHC7 data \cite{TOTEM7DATA} are poorly described in this model, and its modification was proposed that contains a form factor correcting the exponential responsible for small $|t|$. Then, the LHC data at 8 and 13 TeV were described in the same model \cite{Goncalves2019}. However, the region of very small $t$ was excluded from the analysis, and the model gave systematically underestimated description in the large-$t$ region. 

In parallel, the data were described using an amplitude containing three contributions from strong interactions \cite{KFK13,KFK14,KFK21}. One of them was the ordinary Gaussian. The other was inspired by the stochastic vacuum model and provided a slower (Yukawa-type) decrease in the amplitude over long distances. The third term was designed to correct the contributions at large $|t|$ and represented a ``perturbative'' three-gluon exchange \cite{DL1,DL2}. In addition, the Coulomb contributions with Coulomb phase was taken into account in the region of very small $t$. On this basis, a satisfactory description of the ISR and LHC data up to 7 TeV was obtained. The subsequent fit \cite{KFK21} at 13 TeV led to admissible $\chi^2$, but at large $|t|$ the description was overstated compared to the data. On the whole, this description seems admissible since inaccuracies in the large-$|t|$ region can apparently be corrected. However, the physical nature of the first two contributions to the amplitude is still not not quite clear.

In fact, there are quite a few other approaches \cite{Dremin2013}. We have mentioned only those of them, some aspects of which will be used in our study. Along the same row, we also mention \cite{Islam2006,Islam2009}, where the amplitude was described by the sum of three contributions due to the independent scattering of three different layers in the protons. The first contribution was caused by the diffraction process. It was initially described in the impact parameter space as an even combination of Fermi-Dirac distributions. The second contribution was the $\omega$ meson exchange with a form factor providing Orear behavior at large $t$. The third term was determined as a hard Pomeron exchange. All in all, the description was controlled by 17 adjustable parameters. Unfortunately, it strongly disagreed with the LHC data at 7 TeV \cite{TOTEM2011a}. We have marked this model because of its idea of a multilayer interpretation of protons. However, we do not accept the idea that each layer scatters only off a similar layer in another proton, while the other layers remain undetected. We believe that if different layers ``feel'' the presence of other layers in one proton, then they cannot miss them in another proton.

In this paper, we propose an approach which is based on assumption that in the range of small and medium transfers the amplitude can be represented as the sum of three exponentials, each with its own complex phase. (As a first experience, we consider the phases to be constant.) The appearance of three exponentials can be considered as a consequence of scattering by two layers in the protons. Actually, this assumption was put forward earlier on the basis of the observation of two cones in the differential cross section \cite{Orear1978,Dremin2018,Csorgo2019,Dremin2020}. In our approach, it becomes possible to study this phenomenon at the amplitude level. In particular, we will be able to consistently take into account the interference and cross-scattering of the layers. Namely, the exponentials with the maximum and minimum slope will correspond to the scattering of the outer layers off each other and that of the inner layers, respectively. The exponential with intermediate slope will correspond to the scattering of the outer layer off the inner layer and vice versa. Furthermore, we take into account the Coulomb and Coulomb-nuclear contributions in the entire transfer region. In the region of large $|t|$, we introduce the ``perturbative'' three-gluon contribution. Based on this, we describe the data at the ISR and LHC energies and look for changes in the scattering parameters and characteristics with increasing the energy. Ultimately, we study the properties of the layers mentioned above.

The structure of the paper is as follows. In Sect.~\ref{sec2} we define the analytic representation of the amplitude. In Sect.~\ref{sec3} we determine the amplitude parameters based on the ISR and LHC data. Sect.~\ref{sec4} analyses the solutions. In Sect.~\ref{sec5} we discuss the properties of the above layers and a possible cause of their occurrence. Sect.~\ref{sec6} summarizes the results.

\section{Analytic representation of the amplitude}\label{sec2}

Let the elastic amplitude $A(s,t)$ be normalized so that at high energies
\begin{equation}\label{D1}
\frac{\mbox{d} \sigma}{\mbox{d} t} \left[ \frac{\mbox{mb}}{\mbox{GeV}^2} \right] = 
|A|^{2} .
\end{equation} 
Accordingly, the optical theorem is $\sigma_{tot} = 4 \sqrt{\pi} (\hbar c) \times \mbox{Im} A$, where $(\hbar c)^2 \! = 0.389379 $ $ \mbox{mb} \,\mbox{GeV}^2 $. 

The amplitude $A$ is formed due to several contributions of different nature. First of all, this is the contribution due to nonperturbative strong interactions, mainly forming a diffraction pattern of scattering. We call this contribution the nuclear amplitude and describe it as the sum of three exponentials equipped with constant phases:
\begin{equation}\label{D2}
A^N (s,t=-q^2) = 
\mbox{i}\sum_{n=1}^{3} \mbox{\sc A}_n e^{\mbox{\scriptsize{i}} \phi_n} e^{-\mbox{\scriptsize\sc B}_n q^2 /2 }.
\end{equation} 
Here $\mbox{\sc A}_{n}$, $\mbox{\sc B}_{n}$ and $\phi_{n}$ are parameters. In general, they are $s$-dependent, but not $t$-dependent, $\mbox{\sc A}_{n}  \geq 0$, $\mbox{\sc B}_{n} \geq 0$, $-\pi \leq \phi_{n} \leq \pi$, and we assume $\mbox{\sc B}_{1} \geq \mbox{\sc B}_{2} \geq \mbox{\sc B}_{3}$. We emphasize that the above parametrization is not based on Regge theory and does not imply any particular signature of each of the contributions. 

Since protons have electric charge, the Coulomb contributions are significant in the region of very small transfers. In the leading order in the fine-structure constant $\alpha$, the purely Coulomb contribution has the form
\begin{equation}\label{D3}
A^{C} = -Q {\cal F}^2/q^2 \,.
\end{equation} 
Here $Q = 2 \sqrt{\pi} \alpha \, (\hbar c)$ and ${\cal F}$ is the proton form factor. In what follows we use it in the exponential parametrization, ${\cal F}(q^2) = \exp(-2 q^2/\Lambda^2)$, $\Lambda^2 = 0.71 \, \mbox{GeV}^2$.

The Coulomb contributions against the background of strong interactions are studied in \cite{Yennie,Cahn,Kondrat}. We use the result of \cite{Cahn}, namely the intermediate formula (34) of this work. Reducing it to our notation, we write the Coulomb-nuclear amplitude in the form (below, we omit the dependence on $s$)
\begin{equation}\label{D4}
A^{CN}(q^2) = \mbox{i}\alpha \! \int_{0}^{\infty} \!\!
\mbox{d} q'^{\,2} \ln \frac{q'^{\,2}}{q^2} \left[ {\cal F}^2(q'^{\,2}) \widehat{A}^{N}(q'^{\,2}\!,q^2) \right]^{\prime}
\end{equation} 
where the prime after square brackets denotes the derivative with respect to $q'^2$, and
\begin{equation}\label{D5}
\widehat{A}^{N}(q'^{\,2},q^2) = \frac{1}{2\pi}
\int_{0}^{2\pi} \!\!\mbox{d} \phi \, A^{N}(q^2 + 2 q q' \cos\phi + q'^{\,2}).
\end{equation} 
Integral (\ref{D5}) for each term in (\ref{D2}) is calculated explicitly. On this basis, we get without much difficulty
\begin{equation}\label{D6}
A^{CN} = - \alpha \sum_{n=1}^{3} 
\mbox{\sc A}_{n} e^{\mbox{\scriptsize{i}} \phi_n} e^{-\mbox{\scriptsize\sc B}_n q^2 /2 } G_{n}(q^2) \,,
\end{equation} 
\begin{eqnarray}\label{D7}
G_{n}(q^2) &=& 2\ln\left(\frac{\mbox{\sc B}_n q^2}{2}\right) + 2\gamma \\
&& -\mbox{Ei}\left[\frac{\mbox{\sc B}_n q^2}{2}  \left(1 + \frac{8}{\mbox{\sc B}_n \Lambda^2}\right)^{-1}\right] .
\nonumber
\end{eqnarray} 
Here $\gamma$ is the Euler constant, $\gamma = 0.577\dots$, $\mbox{Ei}(x)$ is the exponential integral \cite{Bateman}. 

For small $q^2$ (\ref{D7}) gives 
\begin{equation}\label{D8}
G_{n}(q^2)|_{q^2 \to 0} = \ln\!\left(\!\frac{\mbox{\sc B}_n q^2}{2}\!\right) + \gamma + \ln\!\left(\!1\!+\!\frac{8}{\mbox{\sc B}_n \Lambda^2}\!\right) + \bar{o}(1)\,.
\end{equation} 
This formula reproduces the result \cite{Cahn} for the Coulomb phase with $B$ replaced by $\mbox{\sc B}_n$. However, the factors $1 +\mbox{i}\alpha G_{n}$ appear for each term in (\ref{D2}), and they are not equal to each other. Therefore, these factors cannot be extracted into a common factor at $A^N$, then exponentialized and represented as a common Coulomb phase. This fact distinguishes our approach from \cite{Yennie,Cahn,Kondrat} and, accordingly, from many approaches used in the data analysis.

The next note is related to the fact that exponential integral increases exponentially for $x \to \infty$, $\mbox{Ei}(x) \sim x^{-1} \exp(x)$. As a result, the large-$q^2$ behavior of $A^{CN}(q^2)$ essentially depends on the presence of the form factor ${\cal F}^2(q^{2})$. Really, in the case of its absence, i.e.~at $\Lambda^2 = \infty$, we have $A_{CN}(q^2) \sim 1/q^2$ at $q^2 \to \infty$, while for finite $\Lambda^2$ the  $A^{CN}$ decreases exponentially with a factor $1/q^2$. Nonetheless, the latter decrease is slower than that of $A^{N}(q^2)$. For this reason, we take into account $A^{CN}$ in the entire region~$q^2$.

Let us now turn to the large-$q^2$ region. Here the strong perturbative contributions can appear. At ISR energies their appearance was revealed due to the almost energy-independent behavior of the differential cross section at $|t|\gtrsim 4$ GeV$^2$ of the form
\begin{equation}\label{D9}
\frac{\mbox{d} \sigma}{\mbox{d} t} = 0.09 \, t^{-8} \,.
\end{equation} 
This behavior was explained \cite{DL1,DL2} by three-gluon exchange between colliding protons, with one power of $t^{-2}$ arising from kinematical factor and $t^{-6}$ directly from the three-gluon exchange. Moreover, the corresponding contribution to the amplitude was real and positive in the case of $pp$ scattering. Phenomenologically, it can be described as, cf. \cite{KFK13,KFK14,KFK21,DL3},
\begin{equation}\label{D10}
A^{P} = \kappa \, (\hbar c) \, t^{-4} \left(1 - e^{-a t^6}\right)\,,
\end{equation} 
where $\kappa$ and $a$ are positive parameters. Matching with (\ref{D9}) for large $|t|$ prescribes $\kappa = 0.48$. The last factor in (\ref{D10}) cuts off contributions in the region of small and medium $t$ (in the non-perturbative domain). The requirement for (\ref{D9}) to be established at $|t| \gtrsim 4$ GeV$^2$ implies $a \approx 3 \times 10^{-4}$ GeV$^{-12}$. The parameters $\kappa$ and $a$ can be independently determined if the wave function of fast moving protons is known. Since the size characteristics of the protons change logarithmically with increasing the energy, see e.g.~\cite{Gribov73,Nekrasov1,Nekrasov2}, we expect these parameters to depend weakly on $s$. This is indicated, in particular, by the fact that at $\sqrt{s} = 7$ TeV the power-law behavior of the differential cross section of the form $t^{-8}$ manifests itself starting from $|t| \approx 1.5$ GeV$^2$ \cite{TOTEM2011a}, i.e.~at lower $|t|$ compared to the ISR case.

In fact, the contribution of $A^{P}$ to the main physical characteristics is very small, since in the large-$t$ region the amplitude decreases by several orders of magnitude compared to its values in the peaks region. Therefore, we do not add $A^{P}$ to the definition of $A^{CN}$, which actually is a perturbative contribution. For the same reason, we do not care about the details in the definition of the proton form factor.

Gathering the contributions, we obtain the full amplitude of elastic scattering
\begin{equation}\label{D11}
A(s,t) = A^{N} + A^{C} + A^{CN} + A^{P}\,.
\end{equation} 
It depends on 10 adjustable parameters if $\kappa$ has a fixed value, or on 11 if $\kappa$ is not known a priori.

\section{Determination of amplitude parameters}\label{sec3}

In this section we get a representation in terms of the amplitude (\ref{D11}) of ISR data at 31, 45, 63 GeV \cite{ISR,ISRtables} and LHC TOTEM data at 7 TeV \cite{TOTEM7DATA} and 13 TeV \cite{TOTEM13a,TOTEM13b}. In the ISR case, the choice of energy is determined so that its next value is separated from the previous one by about 1.5 times. In this way, without unnecessary cluttering, we will get an overall picture in the range of ISR energies. In the case of LHC, the choice is determined by the presence of a sufficiently wide measured region of $t$, which would  include areas of very small and large $t$. This condition is satisfied by the TOTEM data. (Further by LHC we mean TOTEM.) In each case when determining the actual data set, we follow the rule \cite{ISR} that when two sets overlap, the data should be taken from the set that has the smallest errors. In all cases except of 7 TeV, we use statistical and systematic errors summed in quadrature. In the particular case of 7 TeV the systematic errors are too large, and we use only statistical errors.

In fact, in each of the selected cases there are several solutions. In order to single out physical ones, we impose the following conditions. First, we demand that the condition $\mbox{Im} A^{N}(s,t=0) > 0$ required by unitarity be satisfied.\footnote{We emphasize that this condition is imposed on the total nuclear amplitude, and not on its individual components corresponding to the internal layers, which themselves do not exist independently of protons.} Second, we demand the positivity of the phase shift of the amplitude $A^{N}$ in the impact parameter representation. Recall, the amplitude in this representation is determined (in our notation) as
\begin{equation}\label{D12}
h(s,b) = \frac{\sqrt{\pi}}{(\hbar c)} \int \frac{\mbox{d}^2 {\bf q}}{(2\pi)^2} \; 
e^{-\mbox{\scriptsize{i}} {\scriptsize \bf q b}} A^{N}(s,t) .
\end{equation} 
In turn, the phase shift $\delta$ is defined in the eikonal parametrization,
\begin{equation}\label{D13}
h(s,b) = 
\frac{1}{2\mbox{i}} \left[\eta(s,b) e^{2\mbox{\scriptsize{i}}\delta(s,b)} - 1\right].
\end{equation}
Here $\eta(s,b)$ is the elasticity parameter, $\eta(s,b) \ge 0$. Thus, 
\begin{equation}\label{D14}
\tan 2 \delta(s,b) = \frac{\mbox{Re} \, h(s,b)}{1/2 - \mbox{Im} \, h(s,b)}\,.
\end{equation} 
The positivity condition for $\delta(s,b)$ follows from the time delay for the passage by colliding particles of the interaction region \cite{Bohm,Messiah}. In view of probable errors at the periphery, we apply this requirement to the internal region of interaction. Practically, we require that the real part of the impact parameter amplitude be positive in this region. This implies $0 < 2 \delta < \pi$. If $\mbox{Re} \, h(s,b)$ turns out to be negative, we additionally require $\mbox{Im} \, h(s,b) > 1/2$ in the same region, which implies $\delta > \pi/2$.\footnote{In fact, the $\delta \approx \pi/2$ case means that the resonant scattering mode is achieved \cite{Nekrasov3}, see also \cite{Anisovich2014,Troshin}. The onset of this mode is possible at asymptotically high energies.} 

\begin{table*}[t]
\caption{The amplitude parameters.}\label{T1}
\vspace*{0.3\baselineskip} 
\scriptsize
\begin{tabular*}{\textwidth}{@{\extracolsep{\fill}}llllllll@{}}
\hline\noalign{\smallskip}
 $\sqrt{s}$ & 31 GeV & 45 GeV & 63 GeV 
 & 7 TeV (I) & 7 TeV (II) & 13 TeV (I) & 13 TeV (II)
\\[1mm]
\hline\noalign{\medskip}
$\mbox{\sc A}_1$  & 8.80$\pm$3.62  & 4.83$\pm$1.52  & 5.67$\pm$2.57  
& 19.7$\pm$3.6 & 20.16$\pm$0.68 & 9.62$\pm$1.03 & 10.85$\pm$1.68 
\\[1mm]
$\mbox{\sc A}_2$  & 6.71$\pm$2.65 & 6.72$\pm$0.17 & 7.30$\pm$0.59 
& 18$\pm$1220 & 30$\pm$5080 & 17.42$\pm$0.61& 17.62$\pm$0.29  
\\[1mm]
$\mbox{\sc A}_3$ & 0.030$\pm$0.003 & 0.027$\pm$0.002 & 0.029$\pm$0.003 
& 15$\pm$1230 & 27$\pm$5080 & 1.70$\pm$0.03 & 1.68$\pm$0.03 
\\[1mm]
$\mbox{\sc B}_1$ & 13.00$\pm$1.34 & 20.04$\pm$2.23 & 19.23$\pm$3.30 
& 21.3$\pm$1.2 & 20.94$\pm$0.42 & 28.44$\pm$0.70 & 27.92$\pm$0.87 
\\[1mm]
$\mbox{\sc B}_2$  & 9.06$\pm$0.54 & 9.78$\pm$0.08 & 10.11$\pm$0.22 
& 7.9$\pm$37.5 & 7.7$\pm$47.8 & 15.64$\pm$0.14 & 15.70$\pm$0.10 
\\[1mm]
$\mbox{\sc B}_3$  & 1.40$\pm$0.10 & 1.36$\pm$0.08 & 1.38$\pm$0.10 
& 7.0$\pm$32.0 & 7.1$\pm$44.0 & 5.18$\pm$0.03 & 5.14$\pm$0.04 
\\[1mm]
$\phi_1$ & 0.76$\pm$0.17 & 0.66$\pm$0.14 & 0.72$\pm$0.16 
& 0.00$\pm$0.54 & 0.10$\pm$0.33 & $-$0.13$\pm$0.75 & 0.30$\pm$0.21 
\\[1mm]
$\phi_2$ & $-$1.17$\pm$0.43 & $-$0.56$\pm$0.25 & $-$0.70$\pm$0.37 
& $-$0.05$\pm$4.78 & $-$0.39$\pm$4.61 & $-$0.03$\pm$0.45 & $-$0.33$\pm$0.17 
\\[1mm]
$\phi_3$ & 1.87$\pm$0.34 & 2.08$\pm$0.26 & 1.94$\pm$0.38 
& $-$3.14$\pm$0.69 & 2.79$\pm$2.55  & $-$2.87$\pm$0.42 & 3.14$\pm$0.15  
\\[1mm]
$\kappa$ & 0.48 (input) & 0.48 (input) & 0.48 (input) 
& 0.23$\pm$0.08 & 0.27$\pm$0.03 & 0.11$\pm$0.04 & 0.14$\pm$0.02 
\\[1mm]
$a$   & (2.9$\pm$0.6)$10^{-4}$ \!\!\!\!\!\!\!& (2.0$\pm$0.4)$10^{-4}$ \!\!\!\!\!\!\!& (2.9$\pm$0.6)$10^{-4}$ \!\!\!\!\!\!\! 
& 0.05$\pm$0.07 & 0.11$\pm$0.05 & 0.018$\pm$0.012 & 0.030$\pm$0.013 
\\[1mm]
$\chi^2\!${\tiny /dof} \!\!\!\!\! & 0.75 & 1.82 & 0.70 
& 2.04 (0.11) & 2.01 (0.11) & 0.96 & 0.98 
\\[1mm]
$N${\tiny data} \!\!\!\!\! & 180 & 228 & 137 
& 165 & 165 & 346 & 346
\\
\noalign{\smallskip}\hline 
\end{tabular*} 
\end{table*} 

\begin{figure*}
\hspace*{-0.06\textwidth}
\includegraphics[width=1.1\textwidth]{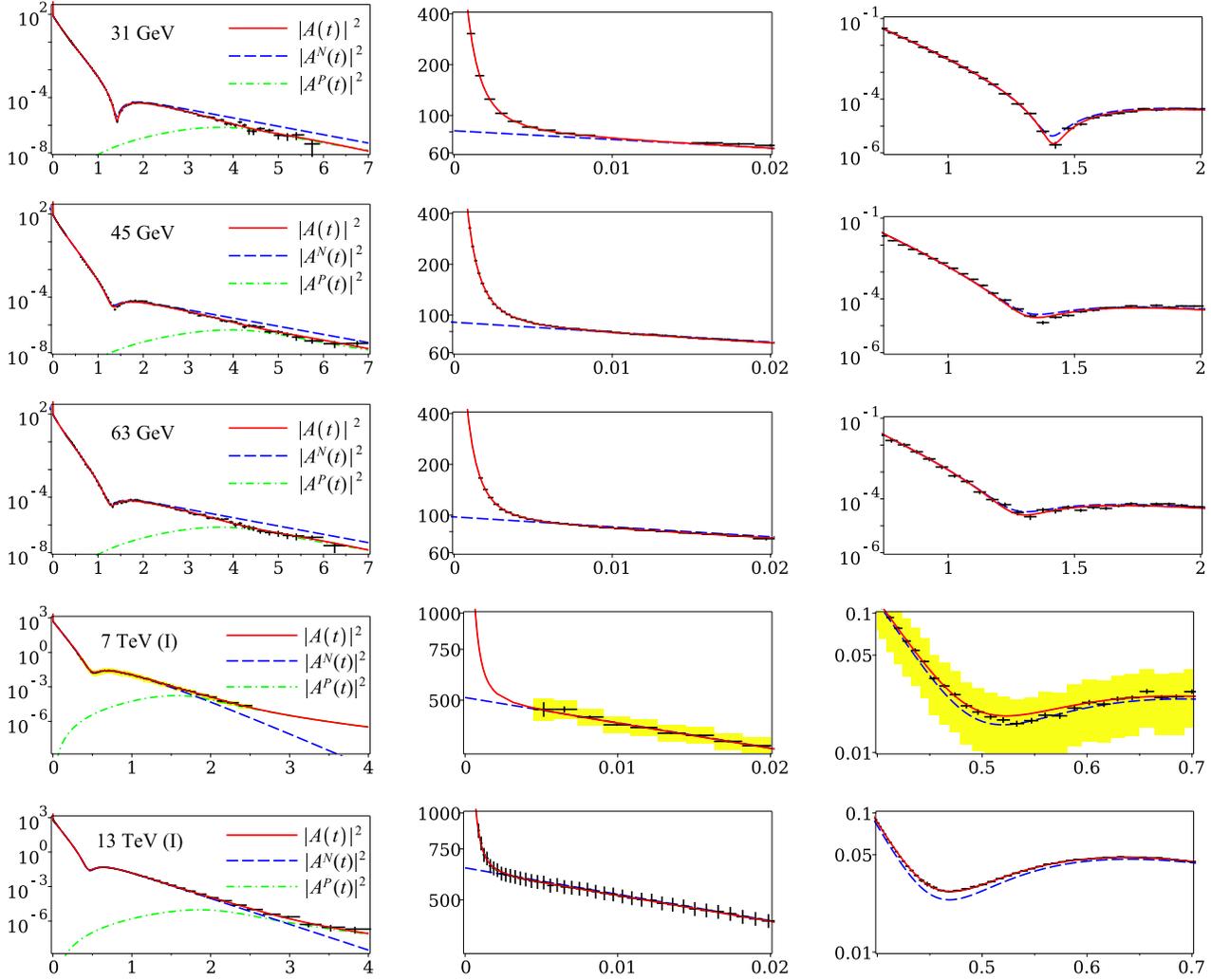}
\caption{\small Differential cross section $\mbox{d} \sigma/\mbox{d} t$ [mb/GeV$^2$] vs.~$|t|$ [GeV$^2$] and some of its components due to strong interactions. Designations are on the left panels. Shown from left to right: the entire region~$t$, the region of very small $t$, and of the dip of the cross section. Everywhere, except 7 TeV, the data are shown with total errors. In the case of 7 TeV, the data are shown with statistical errors, and shaded (yellow) areas show  the total errors.
}\label{Fig1}
\end{figure*}

Further, we expect the behavior of the amplitude and its components to change slowly with the energy. Practically, this means that the behavior of the amplitude should not change significantly when the energy varies within the ISR domain and similarly in the 7--13 TeV region. In some instances in the case of ISR energies, this allows us to discard some solutions (which at the same time have noticeably larger $\chi^2$). In the case of TOTEM energies this condition has no practical consequences. 

Following the above rules, we get one solution in each case at ISR energies and two solutions with almost equal $\chi^2$ in each case at TOTEM energies (in the latter case the solutions I and II actually differ only in the values of $\phi_n$). The results of the fit are presented in Table~\ref{T1}, where $\mbox{\sc A}_n$ are given in $\sqrt{\mbox{mb}/\mbox{GeV}^2}$, $\mbox{\sc B}_n$ in GeV$^{-2}$, $\phi_n$ in radians, and $a$ in GeV$^{-12}$. In each case, we also show $\chi^2$ per degree of freedom and the number of data points used. In the 7 TeV case the $\chi^2$ is determined using statistical errors (the corresponding $\chi^2$ with the total errors is given in parentheses). A comparison of the data with the curves defined with the above parameters is shown in Fig.~\ref{Fig1}. (At the TOTEM energies, the curves for solutions I and II practically coincide; for definiteness solutions of type I are used.) 

\begin{table*}
\caption{The main characteristics of scattering. The Ref.~lines contain the previously obtained values. Our results with higher accuracy are in bold.}\label{T2}
\vspace*{0.3\baselineskip} 
\scriptsize
\begin{tabular*}{\textwidth}{@{\extracolsep{\fill}}llllllll@{}}
\hline\noalign{\smallskip}
 $\sqrt{s}$ & 31 GeV & 45 GeV  & 63 GeV 
 & 7 TeV (I) & 7 TeV (II) & 13 TeV (I) & 13 TeV (II)
\\[1mm]
\hline\noalign{\medskip}
$\sigma_{tot}$ & {\bf 39.85(11)} & {\bf 42.07(7)} & {\bf 43.55(8)} 
& {\bf 99.9(4)} & {\bf 99.8(3)} & {\bf 111.9(3)} & {\bf 112.1(3)} 
\\[1mm]
Ref. & 40.14(17) \cite{ISR} & 41.79(16) \cite{ISR} & 43.32(23) \cite{ISR} 
&\multicolumn{2}{c}{98.0(2.5) \cite{TOTEM7}}
&\multicolumn{2}{c}{110.5(2.4) \cite{TOTEM13a}}  
\\[1mm]
\hline\noalign{\medskip}
$\sigma_{e\ell}$ & 7.15(21) & 7.24(21) & 7.74(34) 
& 25.5(128.0) & 25.5(289.0) & {\bf 31.1(6)} & {\bf 31.2(4)} 
\\[1mm]
Ref. & 7.16(9) \cite{ISR} & 7.17(9) \cite{ISR} & 7.66(11) \cite{ISR}
&\multicolumn{2}{c}{25.1(1.1) \cite{TOTEM7}}
&\multicolumn{2}{c}{31.0(1.7) \cite{Elastic13}} 
\\[1mm]
\hline\noalign{\medskip}
$\rho_0$ & {\bf 0.009(5)} & {\bf 0.059(2)} & {\bf 0.093(3)} 
& {\bf 0.043(57)} & {\bf 0.005(59)} & 0.09(2) & 0.10(2) 
\\[1mm]
Ref. & 0.042(11) \cite{ISR} & 0.062(11) \cite{ISR} & 0.095(11)\cite{ISR} 
&\multicolumn{2}{c}{0.145(91) \cite{TOTEM7}} 
& 0.09(1) \cite{TOTEM13a} & 0.10(1) \cite{TOTEM13a} 
\\[1mm]
\hline\noalign{\medskip}
$B_0$ & 11.8(7) & 13.70(77) & 13.7(1.1) 
& 20.1(50.6) & 20.0(108.1) & 21.14(31) & 21.19(34)
\\[1mm]
Ref. & 12.2(3) \cite{SlopeISR} & 12.8(3) \cite{SlopeISR} & 13.3(3) \cite{SlopeISR} 
&\multicolumn{2}{c}{19.9(3) \cite{TOTEM7DATA}}
&\multicolumn{2}{c}{20.40(1) \cite{TOTEM13b}}
\\
\noalign{\smallskip}\hline 
\end{tabular*} 
\end{table*} 

In all cases except for 7 TeV, the parameters in Table~\ref{T1} are determined with small or reasonable errors. In the exceptional case of 7 TeV the errors are very large for $\mbox{\sc A}_2$, $\mbox{\sc A}_3$ and $\mbox{\sc B}_2$, $\mbox{\sc B}_3$, although the errors are acceptable for other parameters. Therefore, the solutions at 7 TeV are rather arbitrary and should be treated with caution. However, taking into account the correlation matrix, the errors of the main physical quantities turn out to be acceptable in some cases. In this regard, we retain solutions at 7 TeV as optional for further discussion. As for the main physical quantities, we mean those that are usually given when presenting experimental data. They are the total and elastic-scattering cross sections, $\sigma_{tot}$ and $\sigma_{e\ell}$ [mb], the $\rho_0$-parameter which is the ratio of the real to imaginary parts of the nuclear amplitude at zero transfer, and the slope parameter $B_0$ [GeV$^{-2}$] of the diffraction cone at zero transfer. Their values obtained on the basis of the above solutions are presented in Table~\ref{T2} in comparison with those given in the experimental publications. (The quoted uncertainties correspond to 1$\sigma$ of confidence level.) Note that the values we have obtained are close to the refereed ones, and in the cases highlighted in bold they are defined more precisely, despite the presence of two solutions at TOTEM~energies. 

\section{Analysis of the solutions}\label{sec4} 

Let us now find the changes in various characteristics in the transition from ISR to TOTEM energies. From Table~\ref{T1}, it is easy to see four significant differences. The first is that in the ISR case the value of $\mbox{\sc A}_3$ associated with the smallest of $\mbox{\sc B}_n$ is two orders of magnitude smaller than $\mbox{\sc A}_1$ and $\mbox{\sc A}_2$. (At the same time $\mbox{\sc A}_3$ is clearly non-zero given the errors.) In contrast, in the TOTEM case $\mbox{\sc A}_3$ approaches $\mbox{\sc A}_1$ and $\mbox{\sc A}_2$. Second, $\mbox{\sc A}_2$ is close to $\mbox{\sc A}_1$ at ISR energies, but in the TOTEM case $\mbox{\sc A}_2$ dominants with respect to $\mbox{\sc A}_1$. The third difference is that the parameter $\mbox{\sc B}_3$ remains constant at ISR energies and increases sharply when transiting to TOTEM energies.\footnote{In the above cases while referring to TOTEM, we mean solutions at 13 TeV.} In the next section we discuss these features. The fourth difference is a significant decrease in the parameter $\kappa$ in the case of TOTEM and a simultaneous increase in the parameter $a$. This means a weakening of strong perturbative contributions and a decrease in the transfer scale at which they turn on.We associate this behavior with a change in the  wave functions of protons with a significant increase in the energy.

\begin{figure*}
\hspace*{-0.06\textwidth}
\includegraphics[width=1.1\textwidth]{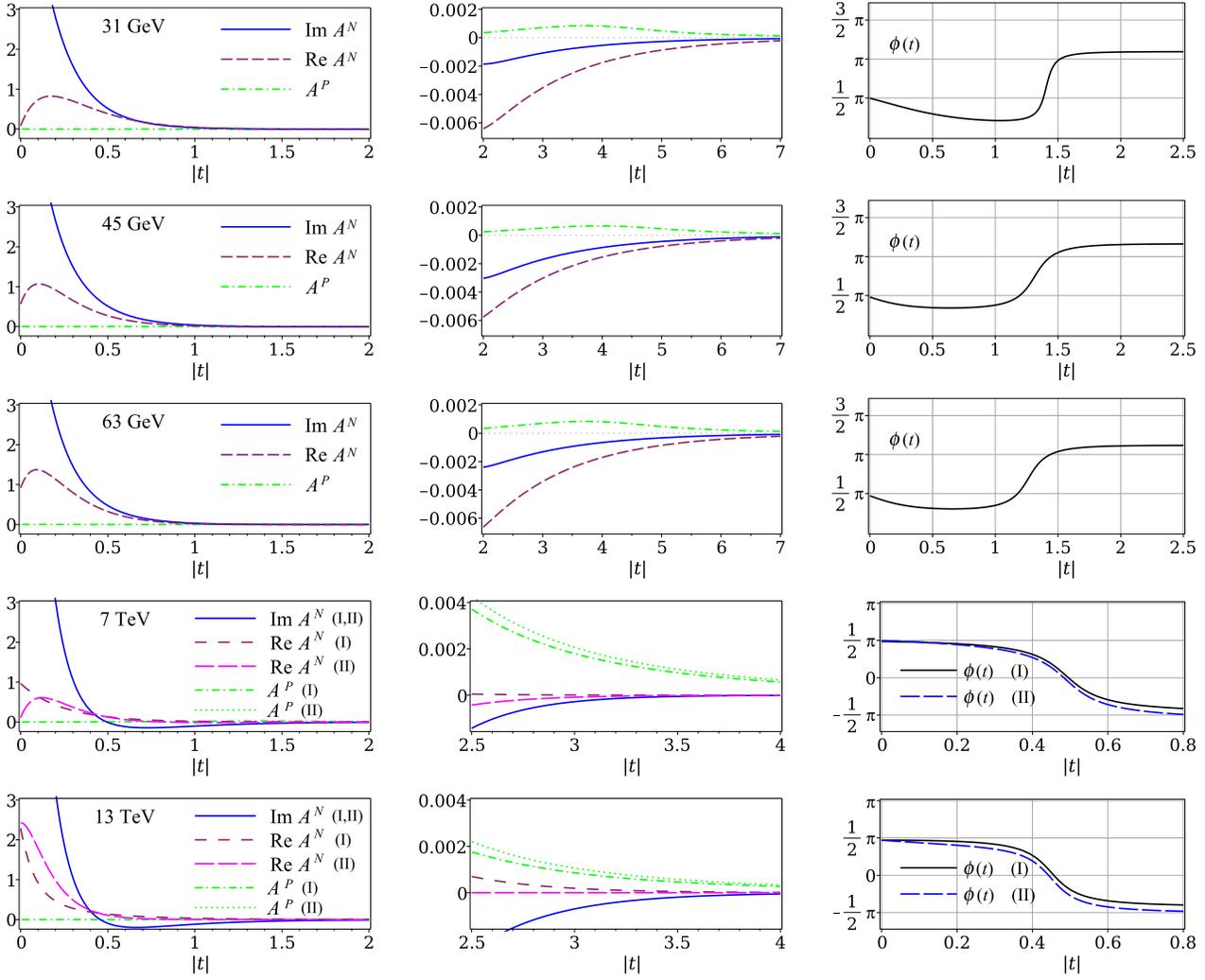}
\vspace*{-\baselineskip}
\caption{\small The contributions to the amplitude due to strong interactions. The regions of small and large $|t|$ are shown in the first two columns. The third column shows the behavior of the complex phase of $A^N(t)$. The corresponding energy is indicated in the first column.}\label{Fig2}
\end{figure*}

Now we turn to the plots in Fig.~\ref{Fig1}. Here we note that the curves generally agree well with the data throughout the whole measured regions of $t$. The next common feature is that $|A^N|^2$ and $|A|^2$ noticeably diverge from each other in the dip region. The difference is about 12\% in TOTEM cases and tens of percent in ISR cases. This means  that the Coulomb-nuclear contributions in all cases are significant in the dip region. Next, we note the difference in the behavior of $|A^N|^2$ and $|A|^2$ in the large-$t$ region. (Here the contributions of $A^C$ and $A^{CN}$ are in fact negligible.) Namely, at ISR energies $|A^N|^2$ exceeds $|A|^2$, while at TOTEM energies $|A^N|^2$ is less than $|A|^2$. To understand why this occurs, consider the plots of individual contributions to the amplitude depicted in the second column in Fig.~\ref{Fig2}. It is easily seen that $\mbox{Re} A^N$ is negative in the large-$t$ region at ISR energies and in absolute value close to the strictly positive $A^P$. As a result, both contributions reduce each other, thereby reducing the full amplitude. At TOTEM energies at large $t$ the contribution of $A^P$ significantly exceeds $|\mbox{Re} A^{N}|$. As a result, there are no noticeable reductions, and $A^P$ is the dominant contribution.

In the region of small $t$, see the first column in Fig.~\ref{Fig2}, $A^N(s,t)$ dominates. A convenient way to give a description of its real and imaginary parts is to consider its total phase, given by the formula
\begin{equation}\label{D15}
A^N(s,t) = |A^N(s,t)| e^{\mbox{\scriptsize{i}} \phi(s,t)}\,.
\end{equation} 
The dependence of $\phi(s,t)$ on $t$ is shown in the third column in Fig.~\ref{Fig2}. In all cases, at $t \to 0$ the $\phi(s,t)$ approaches $\pi/2$ from below (thus ensuring that $\rho_0$ is positive). However, as $|t|$ increases, $\phi(s,t)$ behaves differently at ISR and TOTEM energies. Namely, in the ISR case $\phi(s,t)$ first decreases, then increases, and reaches $\pi/2$ at $|t|=t_R$. Then it reaches $\pi$ at $|t|=t_I$. Here $t_R$ and $t_I$ are the points of the sign change of $\mbox{Re} A^N$ and $\mbox{Im} A^N$, respectively. The values of $t_R$ and $t_I$ are given in Table~\ref{T3}. At TOTEM energies, $\phi(s,t)$ decreases monotonically, crossing $\phi=0$ at $|t|=t_I$, and then crossing or asymptotically approaching the point $\phi=\pi$. Accordingly, $t_R$ is greater than $t_I$ or tends to infinity.\footnote{This behavior fundamentally distinguishes our solutions from the solutions of \cite{KFK13,KFK14,KFK21} and \cite{Pancheri2013} modified in \cite{Pancheri2019}, which are characterized by $t_R < t_I$ behavior.} In Table~\ref{T3}, the infinity symbol denotes a situation when $t_R$ goes beyond the limits of the domain $t$ under consideration. The behavior of the phase in this case is qualitatively similar to the parametrization of Bailly {\it et al.} \cite{Bailly}. A mini-overview of the behavior of $\phi(s,t)$ in various models can be found in \cite{TOTEM8}.

\begin{table*}
\caption{Positions of the dip of $A^N (t)$ and zeroes of $\mbox{Re} A^N$ and $\mbox{Im} A^N$.}\label{T3}
\vspace*{0.3\baselineskip} 
\small
\begin{tabular*}{\textwidth}{@{\extracolsep{\fill}}llllll@{}}
\hline\noalign{\smallskip}
 $\sqrt{s}$ & 31 GeV & 45 GeV 
 & 63 GeV & 7 TeV (I,II) & 13 TeV (I,II)
\\
\hline\noalign{\medskip}
$t_{\mbox{\scriptsize dip}}$  [GeV$^2$]& 1.42 & 1.35 & 1.31 & 0.52 \ \ \  0.52 & 0.47 \ \ \ 0.47
\\[1mm]
$t_{R}\,$ \ [GeV$^2$]& 1.38 & 1.19 & 1.18 & $\infty\:\,$ \ \ \ \ 0.83 & $\infty\enspace\,$ \ \ \ $\infty$
\\[1mm]
$t_{I}$ \ \ [GeV$^2$]& 1.53 & 1.44 & 1.44 & 0.50 \ \ \  0.48 & 0.46 \ \ \ 0.44
\\
\noalign{\smallskip}\hline 
\end{tabular*} 
\end{table*} 

\begin{table*}
\caption{RMS values of $b$ for the total, elasctic, and inelastis scatterings.}\label{T4}
\vspace*{0.3\baselineskip} 
\small
\begin{tabular*}{\textwidth}{@{\extracolsep{\fill}}llllll@{}}
\hline\noalign{\smallskip}
 $\sqrt{s}$ & 31 GeV & 45 GeV 
 & 63 GeV & 7 TeV (I,II) & 13 TeV (I,II)
\\
\hline\noalign{\medskip}
$\sqrt{\langle b^2 \rangle_{tot}}$ [fm]& 0.96(3) & 1.04(3) & 1.04(5) & 1.3(1.6) \ 1.2(3.4) & 1.281(5) \ 1.288(13)
\\[1mm]
$\sqrt{\langle b^2 \rangle_{e\ell}}\:$ [fm]& 0.67(2) & 0.69(2) & 0.70(3) & 0.9(4.3) \ 0.9(9.6) & 0.898(4) \ 0.903(7)
\\[1mm]
$\sqrt{\langle b^2 \rangle_{in}}\,$ [fm]& 1.01(1) & 1.10(1) & 1.11(2) & 1.4(2.8) \ 1.3(6.3) & 1.401(3) \ 1.409(5)
\\
\noalign{\smallskip}\hline 
\end{tabular*} 
\end{table*} 

The above difference in the behavior of the phase has far-reaching consequences. In particular, in view of a theorem by A. Martin \cite{Martin}, which states that the real part of the even signature amplitude cannot have a constant sign near $t = 0$, we can expect that odd contributions, such as the Odderon exchange, become significant at TOTEM energies. This conclusion is consistent with that obtained earlier \cite{Goncalves2019} by comparing data of $pp$- and $\bar{p}p$-scattering at different energies.

The next issue is the behavior of the nuclear amplitude in the impact parameter representation, see definition in (\ref{D12}). The associated object is the overlap function,
\begin{equation}\label{D16}
H_{in}(s,b) = (1 - \eta^2)/4\,,
\end{equation}  
where $\eta(s,b)$ is defined in (\ref{D13}). Both functions, $h(s,b)$ and $H_{in}(s,b)$, are dimensionless and are linked by the unitarity relation, $\mbox{Im} \,h = |h|^2 + H_{in}$. The quantities in this relation are also called profile functions and have the meaning of probability distributions that the total, elastic, inelastic, respectively, scattering occurs at a distance $b$ between the centers of colliding particles. Accordingly, the mean square of the impact parameter $b$ for various types of scattering may be defined as \cite{Kundrat2016}
\begin{equation}\label{D17}
\langle b^2 \rangle_{X} = \frac{\int_{0}^{\infty} b\mbox{d} b \, b^2 D_X(s,b)}{\int_{0}^{\infty} b\mbox{d} b \, D_X(s,b)} \,.
\end{equation} 
Here $D_X(s,b)$ is the appropriate profile function, X = total, elastic, inelastic. The corresponding root-mean-square (RMS) values of $b$ are given in Table~\ref{T4}. 

\begin{figure*}
\includegraphics[width=1\textwidth]{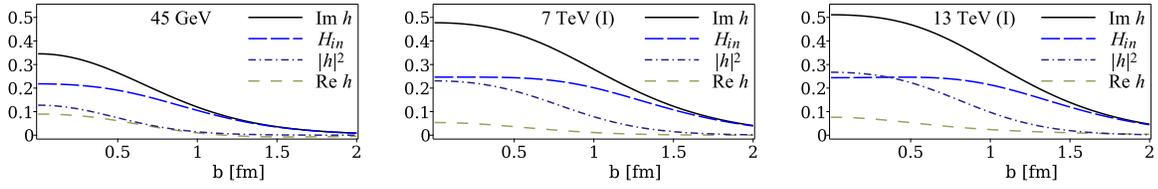}
\caption{\small The profile functions and $\mbox{Re} \,h(s,b)$ at the ISR and TOTEM energies}\label{Fig3}
\end{figure*}

\begin{figure*}
\includegraphics[width=1\textwidth]{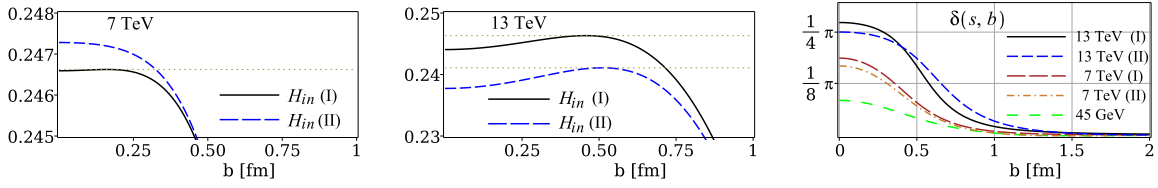}
\caption{\small The left two panels: the overlap function $H_{in}(s,b)$ in higher resolution at 7 TeV and 13 TeV. The right panel shows the phase shift $\delta(s,b)$ at 45 GeV, 7 TeV, and 13 TeV.}\label{Fig4}
\end{figure*}

The plots for the profile functions and for $\mbox{Re} \,h(s,b)$ are given in Fig.~\ref{Fig3}. Due to the very weak energy dependence of these functions, the ISR region is represented only at $\sqrt{s}=45$ GeV. All the functions in the ISR case decrease monotonically as $b$ increases. In the TOTEM case, the differences between the two solutions at 7 and 13 TeV are almost indistinguishable on the chosen scale of the plots. Unfortunately, some details of the behavior of the overlap function $H_{in}$ are not visible in Fig.~\ref{Fig3}. Therefore, we show $H_{in}$ in a higher resolution in the first two panels in Fig.~\ref{Fig4}. At 13 TeV, a small dip in the region of small $b$ becomes clearly visible in the cases of both solutions, approximately within $b<0.5$ fm. Simultaneously, the function $\mbox{Im}\,h$ takes values exceeding the critical value 1/2, and the phase shift $\delta(s,b)$ increases significantly in the vicinity of $b=0$, see the right panel in Fig.~\ref{Fig4}. We interpret the totality of these facts as an indication that the pre-resonant scattering mode is achieved at 13 TeV. (So the resonant scattering mode proper, discussed in \cite{Nekrasov3}, occurs at higher energies.) Previously, a dip in the overlap function at 13 TeV was observed in \cite{KFK21,Troshin2018,Jenkovszky2018,Csorgo2020}. At 7 TeV there is no obvious dip in our approach, but there is a hint of a dip in the case of solution I within $b<0.2$ fm. At the same time, the $\mbox{Im}\,h$ certainly does not reach the critical value 1/2. It is worth noticing that earlier the dip effect in the case of 7 TeV was obtained \cite{Bron2017} by artificially imputing the Bailly phase \cite{Bailly} to the amplitude obtained in \cite{Pancheri2013}.

\section{Transverse structure of fast-moving protons}\label{sec5}

Now consider what we can learn from the above results about the transverse structure of colliding protons. Here the slope parameters $\mbox{\sc B}_n$ are of primary importance. Really, substituting in place of $D_X(s,b)$ in (\ref{D17}) the Fourier-Bessel of the imaginary part of the $n$-th component in $A^{N}$, we get
\begin{equation}\label{D18}
\langle b^2 \rangle_n = 2 \mbox{\sc B}_n\,,
\end{equation} 
where $\langle b^2 \rangle_n$ is the mean-square transverse size of the corresponding interaction region. Recall that the three exponentials in amplitude (\ref{D2}) may be considered as a consequence of scattering of two layers of the protons off each other. Namely, the scattering of the outer layers off each other and the scattering of the inner layers off each other lead to the exponentials with maximum and minimum slopes, respectively, and the scattering of the larger off smaller layers and vice versa leads to the formation of the exponential with intermediate slope. 

Next, we use the fact that, due to the short range of strong interactions, the RMS scattering radius is equal to the sum in quadratures of the radii of the scattering structures \cite{Nekrasov2} (see also discussion in \cite{Petrov-Okorokov}). So, we may write 
\begin{equation}\label{D19}
\langle b^2 \rangle_1 = 2 {\cal R}^2\,,  \quad
\langle b^2 \rangle_2 = {\cal R}^2 + \mbox{\xcm\char'162}^2\,, \quad
\langle b^2 \rangle_3 = 2 \mbox{\xcm\char'162}^2\,,
\end{equation} 
where ${\cal R}$ and $\mbox{\xcm\char'162}$ are the radii of the larger and smaller layers, respectively. A direct consequence of (\ref{D18}) and (\ref{D19}) is a ``sum rule''
\begin{equation}\label{D20}
\mbox{\sc B}_1 + \mbox{\sc B}_3 =  2 \mbox{\sc B}_2\,.
\end{equation} 
In all the cases (\ref{D20}) is fulfilled approximately, which indirectly confirms the picture~under discussion. The values of ${\cal R}$ and $\mbox{\xcm\char'162}$ defined by the first and third relations in (\ref{D19}), as well as the value of the diffraction radius $R_0 = \sqrt{B_0}$, are given in Table~\ref{T5}. 

\begin{table*}
\caption{The diffraction radius and the radii of the transverse layers in colliding protons.}\label{T5}
\vspace*{0.3\baselineskip} 
\small
\begin{tabular*}{\textwidth}{@{\extracolsep{\fill}}llllll@{}}
\hline\noalign{\smallskip}
 $\sqrt{s}$ & 31 GeV & 45 GeV 
 & 63 GeV & 7 TeV (I,II) & 13 TeV (I,II)
\\
\hline\noalign{\medskip}
$R_0$ [fm]& 0.68(2) & 0.73(2) & 0.73(3) & 0.9(1.1) \ \ \ \ 0.9(2.4) & 0.906(7) \ \ 0.907(7)
\\[1mm]
${\cal R}$ \ [fm]& 0.71(4) & 0.88(5) & 0.86(7) & 0.91(2) \ \ \ \ \ 0.902(9) & 1.051(13) \ $\!$1.041(16)
\\[1mm]
$\mbox{\xcm\char'162}$ \ \ [fm]& 0.233(8) & 0.230(6) & 0.232(9) & 0.5(1.2)  \ \ \ \  0.5(1.6) & 0.448(1) \ \ 0.447(2)
\\
\noalign{\smallskip}\hline 
\end{tabular*}
\end{table*} 

It follows from Table~\ref{T5} that while the outer radii $R_0$ and ${\cal R}$ steadily grow with the energy, the inner radius $\mbox{\xcm\char'162}$ remains constant at ISR energies, but increases sharply upon the transition to TOTEM energies. This behavior is unexpected from the  conventional point of view. Indeed, the gradual growth with the energy of the external transverse sizes is predicted in many models, and is generally consistent with the data on the growth of the total and elastic scattering cross sections. However, the size of the inner layers, if it is introduced, usually remains constant independent of the energy. In particular, in \cite{Islam2006,Islam2009} and \cite{Godizov} it is estimated to be a constant of 0.2--0.3 fm, which coincides with $\mbox{\xcm\char'162}$ in Table~\ref{T5} at ISR energies. Below we discuss why the inner radius may start to grow when transiting to TOTEM energies, but first we discuss changes in the activity of the layers in the scattering process.

The mentioned activity can be traced by the change in the values of $\mbox{\sc A}_n$ parameters in relation to the change in the sizes of the corresponding layers. In general, both values should change at the same rate with increasing energy, since both of them squared define the cross section. However, the growth of $\mbox{\sc A}_3$ at the transition to TOTEM energies occurs much faster than the growth of the inner radius. Indeed, according to Table~\ref{T1} and Table~\ref{T5}, the ratio $\mbox{\xcm\char'162}/{\cal R}$ increases by about 1.3--1.6 times, while the $\mbox{\sc A}_3/A_1$ increases by approximately 30--50 times. The difference is huge. This clearly shows that the inner layer becomes much more active in the scattering process. This conclusion is also supported by the fact that $\mbox{\sc A}_2$ noticeably increases with respect to $\mbox{\sc A}_1$ upon transition to 13 TeV.

We emphasize that the above behavior is a consequence of the data and the assumption of a two-layer structure of high-energy protons. So it is of particular interest to identify a model that would explain the above effects. Recall, this is the appearance of the inner layer, the growth of its size, and the increase in its activity in the transition to ultrahigh energies. In fact, such a model exists; it is a modified parton model that takes into account the transverse motions of the partons \cite{Nekrasov1,Nekrasov2}. This model, which is a development of \cite{Gribov73}, predicts a ``diffusive'' growth $\propto\sqrt{\ln s}$ of the external transverse sizes of the hadrons. However, starting from 2--7 TeV, the growth becomes logarithmic. Simultaneously, a rarefaction of partons (a hollow) arises inside the hadrons, which also grows logarithmically in transverse sizes. In fact, the internal rarefaction is a layer, and its growth is consistent with the data in Table~\ref{T5}. 

It is worth discussing the above features in more detail. Actually, at low energies, the above model does not require the rarefaction of partons to occur. Moreover, it assumes by default that the interior of the hadrons is homogeneous. However, this assumption is not mandatory. In particular, in the rest frame, it can be assumed that at the center of the protons there is a decrease in the density of quark matter due to the fact that  the valence quarks are somewhat distanced from each other in order to ensure the tension of the string connecting them. Simultaneously, a dense knot of a Y-shaped flux tube can appear at the center, connecting three valence quarks \cite{Bornyakov}. So, inhomogeneity may well take place at the center of resting protons. Further, in rapidly moving protons, due to time dilation, the role of quantum fluctuations increases and dissociation of valence quarks and gluon structures occurs with the formation of sea quarks and gluons, i.e.~partons. At the initial stage, their motion is characterized by diffuse expansion in transverse directions and the same diffuse in the space of transverse momenta, so that the relation $\Delta R_{\bot} \Delta K_{\bot}\!\sim \ln s$ is satisfied, where $\Delta R_{\bot}$ and $\Delta K_{\bot}$ are the average variances of the distance from the center and transverse momentum of the peripheral partons \cite{Nekrasov1}. Up to complete dissociation, the inhomogeneity within the protons can be preserved. It is quite possible that such an opportunity is realized at ISR energies.

However, when the collision energy further increases and exceeds 2--7 TeV, the behavior of the partons changes radically \cite{Nekrasov1,Nekrasov2}. Namely, instead of diffuse behavior, a mode of correlated motion of the partons is formed, in which the growth of their transverse momenta stops, but the rate of growth of the occupied area increases ($\Delta K_{\bot}\!\sim\!\mbox{const}$, but $\Delta R_{\bot}\!\sim\!\ln s$). The latter effect leads to a decrease in the average density of partons, since the law of growth of the average number of the partons does not change. As a result, the confinement forces pull out the internal partons to the periphery, thus forming a hollow inside.

From this point of view, one can understand why the inner layer becomes more active. The point is that due to a decrease in the local parton density at the center of protons, the effective coupling constant in this region becomes large. For this reason, the inner layer becomes more active when interacting with the partons of another proton involved in the collision. Earlier this phenomenon was discussed in \cite{Nekrasov3}.

\section{ Summary and discussion}\label{sec6}

We have shown that the differential cross section for elastic $pp$ scattering at the ISR and LHC energies can be described using analytical representation of the amplitude as the sum of three exponentials equipped with complex phases, with the addition of the Coulomb and Coulomb-nuclear contributions in the entire transfer region, and strong ``perturbative'' contributions in the large-$t$ region. The proposed amplitude has been considered as a test for the presence of two transverse layers in the scattering protons. The success achieved in the description of the data may be considered as a confirmation of this hypothesis.

Our description is multiparametric, including 10 or 11 parameters in different versions, and it is intended to describe data in a wide range of transfer. In this case, the regions of very small and large $t$ are specifically important, since the real contributions to the amplitude are significant there, which makes it possible to fix the complex phase of the amplitude. So, in order to obtain stable solutions, data covering the outmost areas is needed. This requirement is met by the ISR data and the TOTEM data at 13 TeV and, to a lesser extent, at 7 TeV. Unfortunately, in the latter case some parameters were poorly determined, possibly because of insufficiently wide range of $t$. For the above reason (the difficulty of fixing the complex phase), we did not include in our analysis the LHC data at 8 TeV and lower energies, including ATLAS data, and the available data on the $\bar{p}p$ scattering. 

A distinctive feature of our analysis is that we do not limit the Coulomb and Coulomb-nuclear contributions to the region of very small $t$ and consider them over the entire transfer region. This was justified, since the mentioned contributions turned out to be significant in the region of the dip in the cross section. We believe that this made it possible to improve the accuracy of determining some of the scattering characteristics in comparison with those previously obtained, see Table \ref{T2}.

Among the common properties at all energies, we note a growth with the energy of the external sizes of the protons and the RMS values of the impact parameter for all types of scattering, reconstructed from unitarity. The elastic scattering is central in all cases. However, the inelastic scattering, being central at the ISR energies, becomes peripheral at 13 TeV. (This may be interpreted as reaching the pre-resonant scattering mode at the indicated energy.) Simultaneously, a sharp increase in the size and activity of the inner layer in protons is observed. In fact, the former effect is not surprising. It can be observed as a noticeable increase in the slope of the secondary cone in the differential cross section in the transition to LHC energies \cite{Csorgo2019}.\footnote{Unfortunately, in the case of Gaussian approximation of the cones, the size of the inner layer was mistakenly overestimated in \cite{Csorgo2019} by a factor of two.} At the same time, the effect of a sharp increase in the activity of the inner layer is fundamentally new. We have discussed it and its interpretation in Sect.~\ref{sec5}. 

Another noteworthy change in the transition to LHC energies is a sharp change in the $t$-dependence of the total complex phase of the nuclear amplitude. Moreover, the $t$-dependence of the phase is established in such a way that the real part of the nuclear amplitude does not change sign near $t = 0$. In view of A.Martin's theorem \cite{Martin} this result can be interpreted that contributions with odd signature become significant at ultrahigh energies. There are other changes in the transition to LHC energies, see discussions in Sect.~\ref{sec4}. 

In general, the approach discussed above is not a completed scheme and can be further improved. In particular, one could include the $t$-dependence in the phases $\phi_n$. Of course, this is expedient within a specific model, so as not to increase the number of adjustable parameters. Another direction of modernization is a more accurate account of the Coulomb and Coulomb-nuclear contributions, including second-order contributions in the fine structure constant $\alpha$, see discussion in \cite{Petrov}. In the case of LHC energies, it would be very useful to obtain an independent determination of the parameters of the ``perturbative'' contributions to the amplitude in the region of large $t$. We hope that this will lead to an unambiguous determination of the solutions. Ultimately, this will make it possible to more accurately determine the characteristics of colliding protons at ultrahigh energies.

\begin{flushleft}
{\bf Acknowledgments}
\end{flushleft}

\noindent The author is grateful to V.A.Petrov and A.A.Godizov for useful discussions.

\end{document}